\documentclass{JHEP3} 

\def\ie{{\it i.e.}\ }
\def\eg{{\it e.g.}\ }

\def\etal{{\it et al.}\ }
\def\cf{{\it cf.}\ }

\def\be{\begin{equation}}
\def\ee{\end{equation}}
\def\bea{\begin{eqnarray}}
\def\eea{\end{eqnarray}}
\def\bear{\begin{array}{l}}
\def\eear{\end{array}}

\def\mub{\bar{\mu}}
\newcommand{\D}{\mathcal{D}}
\newcommand{\N}{\mathcal{N}}
\newcommand{\Z}{\mathcal{Z}}
\newcommand{\phibar}{\bar{\phi}}

\newcommand{\phitilde}{\hat{\phi}}
\newcommand{\wh}{\widehat{W}}
\newcommand{\ah}{\widehat{A}}
\newcommand{\nablabar}{\overline{\nabla}}
\newcommand{\mubar}{ \bar{\mu}}
\def\dbar{{d \!\!\bar{\phantom{a}}}\hspace{-.07em}{}^4}
\def\d#1#2#3{\D{#1}(#2,#3)}
\def\di#1{\D{#1}}
\def\dsla{\dbar p \,}

\def\sqr#1#2{{\vcenter{\vbox{\hrule height .#2pt
        \hbox{\vrule width .#2pt height#1pt \kern#1pt
              \vrule width.#2pt}
          \hrule height .#2pt}}}}
\def\Box{\,\sqr{6}{6}}

\def\Lam{\Lambda}
\def\mt{\widetilde{M}}
\def\lam{\lambda}

\def\um{\frac{1}{2}}
\def\inte{\!\int \!}
\def\tr{\,\mathrm{tr} \,}

\def\one{\hbox{1\kern-.8mm l}}

\def\ds{\displaystyle}
\def\ldl{\Lambda \partial_{\Lambda}}

\def\mudmu{\mu \, \partial_\mu}
\def\eq#1{eq.\ (\ref{#1})}
\def\ceq#1{Eq.\ (\ref{#1})}
\def\eqs#1#2{eqs.\ (\ref{#1}, \ref{#2})}
\def\sec#1{sec.\ \ref{#1}}

\def\dphi#1#2{\frac{\delta #1}{\delta \Phi^* {#2}} }

\def\ddphi#1#2#3{\frac{\delta^2 #1}{\delta \Phi^* {#2} \, \delta
\Phi^* {#3} } }
\def\e{\mathrm{e}}

\newcommand{\phihat}{\widehat{\phi}}
\newcommand{\phivec}{{\phi}}
\newcommand{\phistar}{\phi^*}

\newcommand{\gzero}{g_0}
\newcommand{\M}{\mathcal{M}}
\newcommand{\Nuno}{\mathcal{N}=1 }
\newcommand{\Nunostar}{\mathcal{N}=1^* }

\newcommand{\Ndue}{\mathcal{N}=2}
\newcommand{\Nfour}{\mathcal{N}=4}

\title{$\Nunostar$ model and glueball superpotential from
  Renormalization-Group-improved perturbation theory} 

\author{
Stefano Arnone, Francesco Guerrieri and Kensuke Yoshida\\ 
Dipartimento di Fisica, Universit\`a di Roma ``La Sapienza''\\
P.le Aldo Moro, 2 - 00185 ROMA, Italy\\
and\\
 I.\ N.\ F.\ N., Sezione di Roma I\\
E-mails: \email{stefano.arnone@roma1.infn.it, 
francesco.guerrieri@roma1.infn.it, kensuke.yoshida@roma1.infn.it}
}

\preprint{\hepth{0402035}}


\abstract{
A method for computing the low-energy non-perturbative properties of SUSY GFT, 
starting from the microscopic lagrangian model, is presented.
The method relies on 
covariant SUSY Feynman graph techniques, adapted to low energy, and
Renormalization-Group-improved perturbation theory.
We apply the method to calculate the  
glueball superpotential in $\Nuno$ SU(2) SYM and 
obtain a potential of the Veneziano-Yankielowicz type.
}

\begin{document}

\section{Introduction} 
\label{sec:intr}

The study of dualities has %%%produced an incredible amount of work on 
given some unexpected insight into the non-perturbative aspects of  
supersymmetric gauge field 
theories (SGFT), as it was realized that non-per\-turbative effects
at strong coupling can often be captured by some weakly-coupled dual theory.   
The celebrated AdS/CFT correspondence \cite{maldacena} is a clear example 
of the power of such an approach.

Another example is the duality between matrix models (MM) and SGFT
\cite{dijk-vafa}, that is
the fact that the non-perturbative computation of the effective superpotential
in some SGFT reduces 
 to a perturbative calculation in a matrix model whose action is the
 tree-level superpotential.  
%%%although it originates out from the framework of string theory
%%%(``spin-offs'').    
    
The main idea of the MM approach to SGFT 
is just to integrate out the heavy ``matter'' fields
(hypermultiplets) to get the effective glueball
superpotentials, or more precisely the perturbative  corrections to them.

Actually, as was shown in \cite{zanon1}, appealing to the ``duality'' with
the MM can be considered just as a purely technical, efficient
way to calculate the relevant covariant Feynman graphs
\cite{grisaru-zanon1}, corresponding to the low-energy configurations.   

Conservatively \cite{argurio}, one can in fact compute the perturbative
corrections to the leading term in the gluon superpotential - the
Veneziano-Yankielowicz (VY) potential \cite{VY} - adding this latter by hand.

Recently, there have been calculations of the effective superpotential by
means of analogous techniques which  included the leading VY term
\cite{amb,HG}. In \cite{amb}, standard field theoretical manipulations with
super Feynman graphs are employed, which requires the introduction of a
ultraviolet cutoff, while \cite{HG} relies on MM techniques.  

In the latter, one considers a $\Nuno \ U(N_c)$ supersymmetric Yang-Mills
(SYM) theory with an additional chiral superfield in the adjoint
representation of   the gauge group and computes the equivalent MM as
indicated in \cite{dijk-vafa}. The lowest order, \ie the ``quadratic term'' 
in \cite{HG}, reproduces the leading VY potential of the model.

It appears that the VY potential in a pure $\Nuno$ SYM (with no additional
chiral fields) still remains to be computed. This is what we have attempted
in the present note. For the sake of simplicity and clarity, we have
analyzed a SU(2) SYM, the generalization to SU(N) being just a little
cumbersome from a computational point of view.  

Rather than make use of the MM correspondence, as in \cite{HG}, we
rely entirely on the covariant super Feynman graph computations described
in \cite{zanon1}, just as in \cite{amb}. However, given the perturbative
equivalence of the two approaches, our results are closely related to those
of \cite{HG}.

In our piece of work, we have only substituted the physical massive chiral
fields of \cite{amb,HG, Agan} with the auxiliary fields, $( \phi_i,
\bar{\phi}_i)_{i=1}^3$, in the adjoint representation of the gauge group.
 These latter are introduced for the purpose of regularizing the pure
 $\Nuno$ SYM, following the proposal by Arkani-Hamed and Murayama
 \cite{ah-m}, and can be viewed as generalized Pauli-Villars fields.

Further, instead of directly integrating out the massive fields in order to
get the glueball potential, we have applied the technique of the exact
renormalization group (ERG) \cite{Polchinski,Gallavotti,Morris:1994qb}, as
adapted to our particular regularization scheme \cite{afy}.  
 
We vary the regularizing (large) mass, $M_0$,
to a smaller $M$ ($M \ll M_0$, eventually $M \sim 0$) while maintaining
the ``physics'' at energy scales between $M$ and $M_0$ invariant by adding    
suitable (generally non-local) counterterms to the original bare action.
In this way, we compute the ``Wilsonian action'', $S_M$, which is the
solution of the well-known Polchinski's equation \cite{Polchinski} 
with respect to the parameter $M$.

This method has the advantage of guaranteeing the absence of ultra-violet
(UV) divergences 
in our computation and, when necessary, of supplying a ``small'' parameter
for the systematic approximation. 

Naturally, after the ERG transformation has been performed,
we still have to deal with the auxiliary degrees of freedom, which are now,
however, associated to the original $\Nfour$ SYM (\cf \sec{sec:Nuno}) with 
deforming mass $M \sim 0$. The correction they give is not expected to
contribute anything to the purely holomorphic part of the superpotential
\cite{buch}.    

In the end, we can eliminate - at least in the low-energy holomorphic
sector - the auxiliary fields by the well-known limiting process
$M_0 \to \infty, g_0 \to 0$ with the dynamically generated scale,
$\Lambda_{\Nuno}$, 
%%%$$
%%%\Lambda_{N=1} = M_0 \exp -\frac{8 \pi^2}{3 N_c g_0^2}
%%%$$ 
held fixed. 

It is perhaps not too surprising that after the above limit has been taken, the
``residual'' superpotential for pure $\Nuno$  SYM is exactly of the
VY form with its minimum given by
$$
|\bar{s}| = \left| \left \langle \frac{1}{32 \pi^2} W^2 \right \rangle
\right| \sim 
\frac{3}{\rm{e}} \Lambda_{N=1}^3,
$$
a $N_c$-fold solution.

We will leave it to a future publication to make this value more precise
and, in particular, we refrain here from commenting on the famous
controversy between ``weak instanton'' and ``strong instanton'' results
\cite{dorey}.

The paper is organized as follows. In \sec{sec:SUSY} we review the aspects
of the covariant super Feynman graph calculus that will be used in the
following. Then, in \sec{sec:regu}, both our regularization scheme and the
ERG transformation are outlined. The former relies on the finiteness of the
mass-deformed $\Nfour$ SYM \cite{Kovacs:1999rd}, while the latter is an
adaptation of the ERG method for the standard momentum-cutoff-regularized
quantum field theory \cite{Polchinski,Gallavotti,Morris:1994qb}. Sec.\
\ref{sec:Nuno} is devoted to describing in detail the actual calculation of
the glueball superpotential. Finally, in \sec{Conc} we draw some conclusions.

\section{SUSY calculus for low-energy physics}
\label{sec:SUSY}
%%\section{Low energy physics}\label{sec:I}
In the study of non-perturbative properties of SGFT it is possible to
obtain interesting and exact results, such as the evaluation of the
superpotential,  by concentrating on the low-energy, holomorphic, aspects
of the theory. 
In such a ``limited'' domain of application, one is led to expect that some
 basic QFT
techniques (\eg Feynman graphs) can be adapted and formulated in such a
way that one may simplify the computation enough to be able to study
some quantities of interest. 

Such a technique has been put forward in \cite{zanon1} and applied to the
``perturbative'' proof of MM-SUSY QFT correspondence. It consists precisely
in the modification of covariant SUSY Feynman graph techniques of Grisaru
and Zanon \cite{grisaru-zanon1}, adapted to study low-energy physics. 
For the present work the technique allows us to partially replace the
reliance on Konishi anomaly (which was needed, for instance, in
\cite{ah-m,afy}) with a more flexible, and sometimes more precise,
method of computation. In this section, we will give a brief introduction
to the method of \cite{zanon1}. Further details can be found in
\cite{argurio}, beside the original work. We will follow the conventions
introduced in \cite{WB}.

The example we have chosen is the evaluation of the holomorphic part of
the partition function for chiral fields (in the adjoint representation) in
an external gauge field background. 
\be\label{eq:one}
\Z(\mu, \bar{\mu})=\frac{1}{\N} \int \D \phi \D \phibar \exp i\left[ \zeta
  \int d^4 x d^4 \theta \phibar \, e^V \phi +\frac{\mu}{2} \int d^2\theta
  \phi^2 + \frac{\bar{\mu}}{2}\int d^2 \bar{\theta}\phibar^2\right], 
\ee
where  ($\phibar$) $\phi$ is a (anti-)chiral superfield in the adjoint 
representation of the gauge group G=SU(N), and the $1/\N$ takes
care of the trivial UV divergences 
(one may choose for example $\N = \Z(\mu_0, \bar{\mu}_0)$ for
some appropriate $\mu_0$, $\bar{\mu}_0$). %%%[Note by FG we could directly use
%%  $\Z(\mu_0, \bar{\mu}_0)$ instead of $\N$]. 

The first step to evaluate \eq{eq:one} is to integrate out the antichiral
field.  We do this by going over to the gauge chiral representation 
\be\label{eq:chgauge}
\phitilde=\phibar \, \e^V = \e^{-V} \phibar,
\ee
where the last equality holds as $\phibar$ transforms as the adjoint
representation. 
One has
\bea
\overline{\nabla}_{\dot{\alpha}} \phi & = & 0, \nonumber \\
\nabla^{\alpha} \phitilde & = & 0,
\eea
having defined the operators appropriate for the gauge chiral
representation as
\bea
\nabla^{\alpha} & \equiv & \e^{-V}D^{\alpha} \e^{V}, \nonumber\\
\overline{\nabla}_{\dot{\alpha}} & \equiv &
\overline{D}_{\dot{\alpha}}. \nonumber 
\eea
\\
By making use of the generalization
(covariantization) of usual relationships, it is possible to reexpress the
integrals on the chiral or antichiral 
subspace only as integrals on the full Grassmannian space: 
\be\label{eq:cov-relat}
\int d^4x d^2 \bar{\theta} \, \phibar^2 = \int d^4 x d^2\bar{\theta}\,
\phitilde^2=\int d^4x d^2\theta d^2\bar{\theta}\, \phitilde \left(
-\frac{\overline{\nabla}^2}{4\Box_+} \right) \phitilde.
\ee
Now it is possible to ``diagonalize'' the dependence upon $\phi$ and
$\phibar$ in \eq{eq:one} by writing: 
\be
\bear
{\ds \zeta \int d^4 x d^4 \theta \phibar \, e^V \phi +
  \frac{\bar{\mu}}{2}\int d^4 x d^2 \bar{\theta}\phibar^2 }\\[0.35cm] 
{\ds = \int d^4 x d^4 \theta \left[ -\frac{\bar{\mu}}{2}(\phitilde -
    \frac{\zeta}{4 \bar{\mu}}\nabla^2 \phi) \frac{\nablabar^2}{4 \Box_+}
    (\phitilde - \frac{\zeta}{4 \bar{\mu}}\nabla^2 \phi) +
    \frac{\zeta^2}{32 \bar{\mu}}\nabla^2 \phi \frac{\nablabar^2}{4
      \Box_+}\nabla^2 \phi \right]  }\\[0.35cm]
{\ds = \frac{\mubar}{2}\int d^4 x d^2 \bar{\theta}\Big(\phitilde -
  \frac{\zeta}{4 
  \mubar}\nabla^2 \phi\Big)^2 + \frac{
  \zeta^2}{2 \mubar} \int d^4 x d^2 \theta \phi \Box_+ \phi, }
\eear
\ee
where we have used the %%usual 
conventions:
\begin{eqnarray}\label{eq:conventions}
W_\alpha &=& - \frac{1}{4}\bar{D}^2 \e^V D_\alpha \, \e^{-V}, \nonumber \\
\Box_{cov} &=& -\frac{1}{2} \{ \nabla_\alpha, \nablabar_{\dot{\alpha}} \}
\{ \nabla^\alpha, \nablabar^{\dot{\alpha}} \}, %%%= \eta^{\mu \nu}D_\mu D_\nu
\nonumber \\ 
\Box_+&=&\Box_{cov} - W_\alpha \nabla^\alpha - \frac{1}{2}\nabla^\alpha
W_\alpha. 
\end{eqnarray}

So far our transformations of \eq{eq:one} are exact, being only algebraic
manipulations. Now we introduce a series of simplifications valid only for
computations of low-energy physics, such as the determination of the
superpotential. 
In particular, following \cite{zanon1}, we assume that
\begin{description}
\item[i.] $S \equiv W^2$ can be treated as a constant; this of course
  implies that $W^\alpha$ is covariantly constant, \eg $\{ \nabla_\alpha,
  \nablabar_{\dot{\alpha}} \} W^\beta = 0$;
%%%\item[(b)] $\nabla_{\alpha \dot{\alpha}} W^\beta = 0$;
%%%\item[(c)] $W^\alpha$ is covariantly constant;
\item[ii.] the term $\nabla^\alpha W_\alpha$ in \eq{eq:conventions} is
  irrelevant; 
\item[iii.] moreover, moving to the new gauge:
$$
\phi \to \phi' =  \e^V \phi, \qquad
W^\alpha \to W'^\alpha = \e^V W^\alpha \e^{-V},
$$
one has $\phi(\Box_{cov} - W_\alpha \nabla^\alpha)\phi \Rightarrow
\phi'(\Box_{cov} - W'_\alpha D^\alpha)\phi'$; 
\item[iv.] $\Box_{cov}$ can be replaced by the ordinary D'Alembertian, $\Box$.
\end{description}

Under these simplifying assumptions, the partition function \eq{eq:one} is
reduced to  
\be\label{eq:part-func}
\Z(\mu, \mubar) \propto \int \D \phi \exp i \int d^4 x d^2 \theta \left(
\frac{\zeta^2}{2 \mubar}\phi(\Box - W_\alpha D^\alpha) \phi +
\frac{\mu}{2}\phi^2 \right). 
\ee
As is well known, the computation is much easier in momentum space. 
Following the lead in \cite{zanon1}, we will Fourier-transform not
only the ordinary space-time coordinates, but also the Grassmannian ones,
$\theta, \bar{\theta}$: 
$$
\partial_\mu \to - i p_\mu,
$$
$$
D_\alpha \to - i \pi_\alpha,
$$
which will bring about a number of important simplifications. 
\ceq{eq:part-func} can therefore be rewritten as
\be\label{eq:fourier-pf}
%%%\bear
%%%{\ds 
\Z(\mu, \mubar) \propto \int \D \phi^* (p', \pi') \exp \frac{i}{2} 
\int \dsla \,
d^2 \pi \, \phi^*(p, \pi) \bigg( \frac{\zeta^2}{\mubar}( - p^2 + i 
\pi^\alpha W_\alpha ) 
%%%}\\
%%%{\ds
+ \mu \bigg) \phi^*(-p, -\pi),%%% }\\
%%%\eear
\ee
where $\phi^*$ is the Fourier transform of $\phi$.

The ``Feynman rules'' represented in \eq{eq:fourier-pf} exhibit a couple of
very important characteristics which can be used to simplify more complex
computations.

The first is scale invariance: under the rescaling 
$$
p_\mu \to \sqrt{\lambda}p_\mu, \qquad \pi_{1, 2} \to \lambda \pi_{1, 2}
$$
one has $$d^4 p \to \lambda^2 d^4 p, \qquad
d^2 \pi \to \frac{1}{\lambda^2} d^2 \pi$$
\ie the measure $d^4 p  \, d^2 \pi$ is scale invariant.

As a result, one can see that the value of the coefficient $ \zeta^2 / \mub
$ multiplying the momentum part in \eq{eq:fourier-pf} 
is irrelevant. If one chooses, for
instance, $ \zeta^2 /\mub = 1$, then\footnote{Note that this implies 
giving up the information on the antiholomorphic part of the action.}  
\be \label{intbarphi}
\bear
{\ds \Z(\mu, \mub) \Rightarrow Z(\mu) \propto \int \! \d{\phi^*}{p'}{\pi'}
  \exp i 
\int \dsla \, d^2 \pi \, \um \phi^*(p,\pi) \big( -p^2 + i \pi_\alpha W^\alpha
}\\ 
{\ds + \mu \big) \phi^*(-p,-\pi) = \exp -\frac{i}{2} \int \dsla \, d^2 \pi \, 
\tr \ln \big( p^2 + i \pi_\alpha W^\alpha +\mu \big), }
\eear
\ee
where in the second line we have Wick-rotated the momentum $p$ to the
Euclidean one.\footnote{The momentum has been rescaled to rid us of the
  dependence upon $\mub$. Therefore the dimension of $p^2$ is now [mass],
  \ie it is homogeneous with $\mu$.} 

The second concerns the $\pi$ dependence.
The (``one-loop'') integration in \eq{intbarphi} contains the Grassmann
$\int d^2 \pi = 1/2 \int d \pi_1 d \pi_2$, which must be fully absorbed by
the integrand. Thus the only non vanishing contribution comes from the
order-$W^2$ term of the same integrand.
Expanding it in powers of $W \pi$ one has
\be   
\bear
{\ds \Z(\mu) \propto \exp \frac{i}{8} \int \dsla \, d^2 \pi \, 
\tr \Big\{ ( p^2 + \mu)^{-1} (W \pi) ( p^2 + \mu)^{-1} \, (W \pi) \Big\}
}\\[0.35cm] 
{\ds = \exp - \frac{i}{8} \frac{t_2(A)}{16 \pi^2} \int W^2 \int_0^\infty
  \!\!\! d\tau \frac{\tau}{(\tau+\mu)^2 },  }
\eear
\ee
where $t_2(A)$ is the Dynkin index of the adjoint representation of the
gauge group [$t_2(A) = N_c$ for SU$(N_c)$] and $\tau = p^2$.

The above expression is divergent. On the other hand, one can compute 
\be \label{mudmu}
\bear
{\ds \mudmu \ln \Z(\mu) = \left\langle i \frac {\mu}{2} \int \!\! \phi^2
\right \rangle = \frac{i}{8} \frac{t_2(A)}{16 \pi^2} \! \int \! W^2 \!\!
\int_0^\infty \!\!\!d\tau \frac{-2 \tau \mu}{(\tau+\mu)^3 }  }
{\ds = - \frac{i}{16} \frac{t_2(A)}{8 \pi^2} \!\! \int\!\! W^2. }
\eear
\ee

Integrating \eq{mudmu} one obtains
\be \label{Zratio}
\frac{\Z(\mu_2)}{\Z(\mu_1)} = \exp - \frac{i}{16} \frac{t_2(A)}{8 \pi^2} \int
\ln \left( \frac{\mu_2}{\mu_1} \right) W^2.
\ee

The expression \eq{Zratio} shows its close connection with the Konishi
anomaly~\cite{konishi}. As a matter of fact, if one rescales the $\phi$
fields in \eq{intbarphi} by $\lambda$, \ie letting $\phi^*(p,\pi) 
\rightarrow \lam
\phi^*(p,\pi)$ and takes into account the (possibly non-trivial)
corresponding Jacobian,
\be
 \int \! \di{\lam \, \phi^*} = J(\lam) \int \! \di{\phi^*}, 
\ee
then 
\be
\Z(\mu) = J(\lam) \int \! \di{\phi^*} \exp i
\int \dsla d^2 \pi \um \phi^* \Big[ \lam^2 \big( -p^2 + i \pi W \big) +
  \lam^2 \mu \Big] \phi^*.
\ee
Now, exploiting the scale invariance of the measure $\dsla \, d^2 \pi$, 
the $\lam$ multiplying the momentum part can be set to one, leaving
\be
\Z(\mu) = J(\lam) \int \! \di{\phi^*} \exp i
\int \dsla d^2 \pi \um \phi^* \Big[ -p^2 + i \pi W +
  \lam^2 \mu \Big] \phi^* = J(\lam) \Z(\lam^2 \mu).
\ee
Thus $J(\lam) = \Z(\mu)/ \Z(\lam^2 \mu) = \exp \frac{i}{8} \frac{t_2(A)}{8
  \pi^2} \int (\ln \lam) W^2$, which is indeed the correct
value~\cite{konishi}. 

The computation illustrated above is equivalent to the old-fashioned
Feynman graph method of determining the anomalies in SYM 
model~\cite{Gates:1983nr,afy}. The method proposed
in~\cite{zanon1}, 
instead, recognizes the common principles in those approximate computations 
and reformulates them as a method for efficiently extracting the
holomorphic part of the superpotential. 

\section{Regularized $\Nuno$ model}
\label{sec:regu}

\subsection{$\Nunostar$ model}
\label{subsec:nuno}

In the absence of a general, symmetry-preserving cut-off scheme 
for supersymmetric gauge field theories \cite{TM1}, 
we adopt the following regularization, applicable only to the
limited class of SGFT models originally suggested by Arkani-Hamed and
Murayama \cite{ah-m}.

We make use of the fact that four dimensional $\Nfour$ SYM is UV finite
\cite{Sohnius:1981sn,Kovacs:1999rd}. The
classical action for the model is 
\be \label{esseN4}
\bear
{\ds S_{\mathcal{N}=4} (V,\phi_i,\phibar_i; g_0) = \frac{1}{16} \inte d^4 x \, d^2
\theta \frac{1}{\hat{g}_0^2} W^a_{\alpha} W^{a \alpha} +
\inte dx \, d^4 \theta  \left( \frac{2}{g_0^2}\right) t_2(A)
\times}\\[0.3cm] 
{\ds \sum_{i=1}^3 \phibar_i e^V \phi_i 
+ \inte d^4 x \, d^2 \theta \, \left( \frac{\sqrt{2}}{2 g_0^2} 
\right) i f_{abc} \, \phi^i_a \, \phi^j_b \, \phi^k_c
\, \frac{\epsilon^{ijk}}{3!} + h.c., }\\
\eear
\ee 
all the relevant fields transforming as the adjoint representation of the
gauge group.
\ceq{esseN4} is written in the so-called holomorphic form, which is
equivalent to the more usual canonical form with the rescaling  
$W^a(V) \rightarrow W^a(g_c V_c)$. (However one must pay attention to the
corresponding Konishi anomaly \cite{konishi,ah-m}. See appendix A.)
The holomorphic gauge coupling constant, $\hat{g}_0$, is given by
\be \label{eq:3.2}
\frac{1}{\hat{g}_0^2} = \frac{1}{g_0^2} + i \frac{\theta_0}{8 \pi^2}.
\ee

Now it is known \cite{Kovacs:1999rd} that the quantization of the model with
classical action  $S_{\mathcal{N}=4}$ is still free from UV divergences
even in the presence of mass deformations, \ie terms of the form
\be \label{massdef}
\frac{1}{2} \sum_{i=1}^3 \inte d^2 \theta \, M_{0i} \, \phi_i^2 + h.c.
\ee
Moreover, it is believed to be also free of infra-red (IR) divergences if
all the external states are gauge invariant \cite{Kovacs:1999rd}.

Thus, we assume that the partition function
\be \label{part}
\bear
{\ds \Z_{M_0} (\mbox{sources}) = \inte \di{V} \inte \prod_i \di{\phi_i} \, 
\di{\phibar_i} \exp i \Big[ S_{\mathcal{N}=4} (V,\phi_i,\phibar_i; g_0) 
}\\
{\ds + \frac{1}{2} \sum_{i=1}^3 \inte M_{0i} \, \phi_i^2 + h.c. +
  \mbox{gauge invariant sources} \Big]  }
\eear
\ee
be well defined for an arbitrary set of masses $\vec{M}_0 =
\left(M_{0i}\right)_{i=1}^3$. 

By choosing the special case,
\be
\vec{M}_0 = (M_0,M_0,M_0), \label{nuno}\\
%%%\vec{M}_0 &=& (M_0,M_0,0), \label{ndue}\\
%%%\nonumber\eea
\ee
one can realize the regularized model which, at energy scales much lower
than $M_0$, gives the physics of $\Nuno$ SYM.

In the limit that $M_0 \to \infty$ and $g_0 \to 0$ with
\be
\Lam_{\Nuno} = M_0 \exp - \frac{8 \pi^2}{3 N_c g_0^2}, \label{lamnuno}\\ 
%%%\Lam_{\Ndue} &=& M_0 \exp - \frac{4 \pi^2}{N_c g_0^2} \label{lamndue}\\
%%%\nonumber\eea
\ee
held fixed, we have pure $\Nuno$ SYM.

\subsection{Renormalization group transformation}
\label{subsec:reno}

As has been suggested in \cite{ah-m,afy}, we can compute the effective
``Wilsonian'' action, $S_M$, by varying the regularizing mass, $M_0$ (which
is to be much bigger than any physical scale we are interested in) to a
much smaller value, $M$, while keeping the physics (that is the numerical
value of $\Z_{M_0}$) unchanged. We look for the transformation which implements
the equivalence relation:
\be \label{equiv}
\bear
{\ds \Z_{M_0} = \inte \di{V} \!\!\inte\! \di{\phi_i} \, 
\di{\phibar_i} \exp i \Big[ S_{\mathcal{N}=4} (V,\phi_i,\phibar_i; g_0) 
+ \frac{M_0}{2} \sum_{i=1}^3 \inte \phi_i^2 + h.c. +
  \mbox{sources} \Big]  }\\
{\ds = \inte \di{V} \!\!\inte\! \di{\phi_i} \, 
\di{\phibar_i} \exp i \Big[ S_M (V,\phi_i,\phibar_i) 
+ \frac{M}{2} \sum_{i=1}^3 \inte \phi_i^2 + h.c. +
  \mbox{ren. sources} \Big]. }
\eear
\ee
%%%where $\mu=3 (2)$ for $\Nuno (2)$ model.

In general, $S_M$ is non-local and expressed as a functional integral over
some auxiliary fields. %%%, $\phi_i', \phibar_i'$. 
This is the adaptation of the
so-called Exact Renormalization Group (ERG) method for the usual
momentum-cutoff-regulated quantum field
theory \cite{Polchinski,Gallavotti,Morris:1994qb}, where the equality
\be
\bear  
{\ds \int_{0<|p|<\Lam_0} \di{\Phi^*}(p) \, \exp \big( - S'_{\Lam_0}(\Phi)+
  \mbox{sources} \big) }\nonumber\\
{\ds = \int_{0<|p|<\Lam} \di{\Phi^*}(p) \, \exp \big( - S'_{\Lam}(\Phi) +
  \mbox{ren. sources} \big) }\\
\eear
\ee
is required. In the above, $S'_{\Lam_0}(\Phi)$ is the bare action with
UV cutoff $\Lam_0$.

In some simple models, we can implement the condition $0<|p|<\Lam
  (\Lam_0)$ at the level of the propagator by singling out the regulated
  kinetic term from the action, \ie
\be
 S'_{\Lam}(\Phi) = \um \int \Phi^*(-p) \frac{D^{-1}(p)}{K_\Lam(p)}
  \Phi^*(p) + S_{\Lam}(\Phi), 
\ee
where $K_\Lam(p)$ is the cutoff profile corresponding to $0<|p|<\Lam$ and
  again $\Phi^*$ is the Fourier-transformed field. 

$S_\Lam$ satisfies the well-known Polchinski's equation \cite{Polchinski}
\be \label{poeq}
\ldl S_\Lam(\Phi) = - \frac{1}{2} \int \dsla \, D(p) \, \ldl K_\Lam \left(
\ddphi{S_\Lam}{(p)}{(-p)} - \dphi{S_\Lam}{(p)} \dphi{S_\Lam}{(-p)} \right).
\ee

One can easily see, then, that in the limit that $\Lam \to 0$, $S_\Lam$
contains all the information about the complete solution of the original
model with ultraviolet cutoff $\Lam_0$, \ie 
\be \label{limit}
\lim_{\Lam \to 0} S_\Lam (\Phi) = W_{\Lam_0} (J), 
\ee
with $ W_{\Lam_0} (J)$ being the generating functional of
connected Green's functions with UV cutoff $\Lam_0$ \cite{Morris:1994qb}.

As has been shown in \cite{afy}, our action $S_M(V,\phi_i, \phibar_i)$ in
\eq{equiv} satisfies the analogue of \eq{poeq} with $\Lam$ replaced by $M$
when the contribution of the Konishi anomaly is subtracted \cite{afy}.
 
From a physical point of view, we may assume that the analogue of
\eq{limit} is also valid for our simplified regularization and
corresponding RG transformation, \eq{equiv}. Therefore we assume that for
small enough $M$ ($M \ll M_0$), the action $S_M(V,\phi_i, \phibar_i)$ in
\eq{equiv}, \ie the solution to the anomaly-corrected Polchinski's
equation, describes with good approximation the physics at energy scales
$M \le E \ll M_0$, with no further quantization procedure (path integral).  

We can compute $S_M$ by generalising the
Zinn-Justin's transformation \cite{Zinn-Justin:1992jq,Bonini:1995kp},
originally in the form: 
\be \label{ZJ}
\bear
{\ds \int \di{\Phi} \exp \left[ -\um \inte \Phi^*(-p) \frac{1}{D_1(p) + D_2(p)}
\Phi^*(p) - V(\Phi) \right] }\\[0.35cm]
{\ds = \int \di{\Phi} \, \di{\Phi'} \exp \bigg[ -\um \inte \Phi^*(-p)
    D_1^{-1}(p) \Phi^*(p) -\um \inte \Phi^{'*}(-p)
    D_2^{-1}(p) \Phi^{'*}(p) }\\[0.35cm]
{\ds - V(\Phi+\Phi') \bigg] }
\eear\ee
%%%with for instance, $p^2 D_1(p) = K_{\Lam'}(p) - K_{\Lam}(p)$, $p^2
%%%D_2(p) = K_{\Lam_0}(p) - K_{\Lam'}(p), \Lam<\Lam'<\Lam_0$, 
to the much simpler form of Gaussian integration:
\be \label{simpleZJ}
\exp i \frac{M_0}{2} \inte \phi_i^2 = \frac{\ds
\int \di{\phi'_i} \, \exp i \inte
\Big[  \frac{M}{2} (\phi_i - \phi'_i)^2 + \frac{\mt}{2}
  \phi_i^{'2} \Big] 
}
{\ds
\int \di{\phi'_i} \, \exp i \frac{M+\mt}{2} \inte \phi_i^{'2} 
},
\ee
where $\mt$ is the ``reduced'' mass, defined by $1/M_0 = 1/M + 1/\mt$.

Inserting \eq{simpleZJ} in the first line of \eq{equiv} yields the 
formal expression for $S_M$:
\be \label{SM}
\exp i S_M %(V,\phi_i,\phibar_i) 
= \frac{\ds
\inte\! \di{\phi'_i} \, 
\di{\phibar'_i} \exp i \Big[ S_{\mathcal{N}=4} (\cdots,
  (\phi'_i,\phibar'_i)_1^3;g_0) 
+ \frac{M}{2} \sum_{i=1}^3 \inte (\phi_i-\phi'_i)^2 + h.c. \Big] 
}
{\ds 
\inte\! \di{\phi'_i} \, 
\di{\phibar'_i} \exp i \frac{M+\mt}{2} \sum_{i=1}^3 \inte \phi_i^{'2} + h.c. 
}. 
\ee 

\section{$\Nuno$ Super Yang-Mills and glueball superpotential}\label{sec:Nuno}

The partition function for the regularized $\Nuno$ SYM is given by 
\eq{part} with the mass configuration, \eq{nuno}. 

One has to deal with all the three  
auxiliary fields $ ( {\phi_i} )_{i=1}^3 $ which enter the bare action
$S_{\Nfour}$ with a cubic interaction term. 
%%%However, due to the presence of
%%%the completely antisymmetric tensors $\epsilon^{ijk}, f_{abc}$, the original
%%%action is still quadratic in any pair of auxilary fields (\cf
%%%\eq{esseN4}).
As the original action is still quadratic in any pair of auxilary fields (\cf
\eq{esseN4}) \cite{dijk-vafa}, in order to compute the Wilsonian action
$S_M$, one can proceed in two 
stages. In the first stage, one integrates out two of the three
massive fields - say $\phi_1, \phi_2$ - by means of the RG transformation
outlined in 
\sec{sec:SUSY}. In the second stage, one moves on to integrating out (\ie
reducing by RG procedure) the remaining chiral field, $\phi_3$.\footnote{A
  similar idea  has been used in \cite{zanon1}.}

In what follows, in order to ease notation, 
$\phi_3$ ($\phibar_3$) will be replaced by $\phi$ ($\phibar$).

\subsection{${\phi_1}$ and ${\phi_2}$ reduction}
\label{sec:phi1phi2}

One applies the process described in the previous section keeping
${\phi}$ out as if it were an external field. 

Following \sec{sec:SUSY}, we integrate out  
$\phihat_i = \phibar_i \,e^V, i = 1, 2.$  Then $\Z_{M_0}$ is
reduced to  
\be\label{eq:5.2} 
\bear  
{\ds \Z_{M_0} = \int \di{V} \di{\phi} \di{\phibar} \prod_{i=1}^ {2} \di{\phi_i}
\di{\phihat_i}
\exp i \bigg[ \frac{1}{2} \inte \dsla d^2 \pi \phistar_{ia}(p,
  \pi) \Big( - p^2 + \pi \widehat{W} }\\[0.35cm]
{\ds  + \widehat{A} + M_0\Big)_{(ia,
jb)} \phistar_{jb}(-p, -\pi) + \frac{M_0}{2} \inte \sum_{i=1}^2
\phihat_i^2 + \frac{1}{\gzero^2} \phibar_{a} (e^V)_{(ab)}
\phi_{b} }\\[0.35cm]  
{\ds + \frac{M_0}{2} \inte \phi^2 + \frac{\overline{M}_0}{2}
\int \phihat^2 + \mbox{ (V sources)} \bigg].}  
\eear
\ee  
%%%where $i, j = 1, 2 \quad a, b = adj (\mbox{SU}(N_c)).$
The matrix $\ah(\phi)$ in the above corresponds to the cubic interaction in
$S_{\Nfour}$ and it is given by
\be \label{ah}
\ah(\phi) = \frac{1}{\sqrt{2} g_0^2} (\vec{\phi} \cdot \vec{F})_{ab}\,
\epsilon_{ij},
\ee 
 with $F^a$ being the generators of  the adjoint representation of
 SU($N_c$), $a,b$ the corresponding indices and $i,j=1,2$.  
Also $\wh(\phi) = (i \vec{W} \cdot \vec{F})_{ab} \, \delta_{ij}$.

Now \eq{eq:5.2} is Gaussian in the auxiliary fields, $\phi_i^*(p,\pi)$
[which, in this approximation, are decoupled from the antiholomorphic
  $\phitilde_i$]. Hence one may be tempted to integrate out the auxiliary
fields directly so as to obtain the (low-energy part of the) effective
action.
 
However, there are two obstacles to this line of reasoning.\\
First of all, the integrals over the (Euclidean) 4-momentum are divergent.\\
Secondly and more importantly, perhaps, 
in some computations \cite{afy2} it is useful to have a small
parameter through which to attempt a systematic approximation. Indeed, the
parameters with dimension of mass are $M_0$ and $\phi$. %%\sim \langle \phi
%%\rangle$.  
%%%$M_0$ is expected to be large but in the weak coupling region -
%%%where one can hope to get concrete results - $|\langle \phi
%%%\rangle|$ is large too. 
$M_0$ is expected to be large and $\phi$, being integrated over, can be
large as well.
As a consequence, there is no certainty about the
order of magnitude of, \eg, the ratio $\phi / M_0$.

It would be much more convenient if one could introduce some definitely
small parameter, $M$, such that $M/|\phi| \ll 1 (M\ll M_0)$. The ERG
approach outlined in \sec{subsec:reno} is precisely the answer to
this problem, as it amounts to lowering the
regularising mass without changing the physics.       
The price to pay for such a strategy, though, is that one does not obtain
immediately the effective action, but the ``Wilsonian''
$S_M(V,\phi,\phibar,\phi_i, \phibar_i) - S_{\Nfour}$ [\cf \eqs{equiv}{SM}].
%%%with the hope that the holomorphic corrections due to the integration  
%%%of the original $S_{\Nfour}$ over the auxiliary fields are not significant
%%%\cite{buch}. 
Holomorphic corrections due to the integration of the original $S_{\Nfour}$
over the auxiliary fields are not expected to be significant in our
case. In fact, as shown by Dorey \etal \cite{buch}, these contributions 
are proportional to the product $M_1M_2M_3=M^3$, and therefore vanish in
the limit that $M\to 0$.

Applying the RG transformation of \sec{sec:regu} (\ie $M_0 \to M \ll
M_0$), \eq{eq:5.2} takes the form 
\be\label{eq:5.3}
\bear
{\ds \Z_{M_0} = \inte \di{V} \di{\phi} \di{\phibar} \prod_{i=1}^{2} \di{\phi_i}
\di{\phihat_i} \prod_{i=1}^{2} \di{\phi'_i}
\exp i \bigg[ \frac{1}{2} \inte \dsla d^2 \pi %%%}\\ [0.40cm]
%%%{\ds 
\phistar_{ia} \Big(- p^2 + \pi \wh + \ah 
} \\[0.40cm] 
{\ds + M\Big)_{(ia, jb)}
  \phistar_{jb} + {\phistar_{ia}}' \Big \{
\Big(- p^2 + \pi \widehat{W} + \widehat{A} + M_0\Big)
\Big(- p^2 + \pi \widehat{W} + \widehat{A} + M\Big)^{-1}\Big\}_{(ia, jb)}
  {\phistar_{jb}}' } \\[0.40cm] 
{\ds + \frac{\overline{M}_0}{2} \int \sum_{i=1}^2 \phihat_i^2 + \cdots
\bigg], } 
\eear
\ee
where the ellipsis refers to terms depending upon $V,\phi, \phibar$ and to
source terms. 
[The particular form of the propagators in \eq{eq:5.3} results from
diagonalising $\phi_i, \phi_i'$ in \eq{SM}.]

The $( \phi_i, \phihat_i)$ part of the path integral should reproduce the
relevant part 
of the original bare action $S_{\Nfour}$, with
%%%(V, \phi_i, \phibar_i;\gzero)$, with 
altered mass M, while
the integration over $ {\phivec'_{1, 2}}$ should contribute 
to the non-trivial part of the Wilson action $S_M - S_{\Nfour}$.

The Gaussian integral over $ {\phivec'_{1, 2}}$ gives a
convergent integral over the  
4-momentum. The relevant part of $\Z_M$ depending on $ {\phivec'_{1, 2}}$
becomes: 
$$
\exp - \frac{i}{2} \int \dbar p d^2 \pi \tr \Big\{ \ln\big( p^2 + \pi
\widehat{W}  
+ \widehat{A} + M_0\big) 
- \ln \big( p^2 +  \pi \widehat{W} + \widehat{A} + M\big) \Big\} , 
$$
and, after integration over $\pi_1, \pi_2$,
\be\label{eq:5.4}
\bear
{\ds \exp -\frac{i}{8} \int \dbar p \tr \bigg\{ \bigg[ (p^2 + M_0)^{-2}(1 + 
\frac{\widehat{A}}{p^2 + M_0})^{-1}
\wh_1 }\\[0.40cm]
{\ds (1 + \frac{\widehat{A}}{p^2 + M_0})^{-1} \wh_2 -
  (\wh_1 \leftrightarrow \wh_2) \bigg]  } 
\\[0.40cm]
{\ds - \bigg[ (p^2 + M)^{-2}(1 + \frac{\widehat{A}}{p^2 + M})^{-1} \wh_1 
(1 + \frac{\widehat{A}}{p^2 + M_0} )^{-1} \wh_2 - (\wh_1
    \leftrightarrow \wh_2) \bigg] \bigg\}, } 
\eear
\ee
where the 4-momentum $p_\mu$ has been Wick-rotated.

From now on, for the sake of definiteness and simplicity, we limit 
ourselves  to the
SU(2) gauge group.\footnote{Refer to \cite{afy2} for the generalization to
  the SU($N_c$) case.} 
Then the matrices $\widehat{A}$ and $\widehat{W}$ are explicitly:
\bea
\widehat{A}_{(ia, jb)} & = & i \phi_c \, \epsilon_{cab}\, \epsilon_{ij}, \\
\widehat{W}_{\alpha (ia, jb)} & = & i (W_\alpha)_c \, \epsilon_{cab}\, \delta_{ij},
\eea
 with $ \alpha = 1, 2 \quad i, j = 1, 2 \quad a, b, c = \mbox{ adj (
   SU(2)) } = 1, 2, 3.$ 

In \eq{eq:5.4} we have to perform the integration of a function of the
form: 
\be\label{eq:5.7}
\bear
{\ds \frac{1}{p^2 + \mu^2} \tr \left \{ \Big( 1 + \frac{\widehat{A}}{p^2 + \mu^2}
  \Big)^{-1}  
\widehat{W}_1  \Big( 1 + \frac{\widehat{A}}{(p^2 + \mu)^2} \Big)^{-1}
\widehat{W}_2 \right\} }\\[0.35cm]
{\ds = 4 \left(1 + \frac{\phi^2}{(p^2 + \mu)^2} \right)^{-2} \left\{ W_1 W_2 - 
2 \frac{(\phi \cdot W_1)(\phi \cdot W_2)}{(p^2 + \mu)^2} + \frac{\phi^2(W_1
  W_2)}{(p^2 + \mu)^2}  
\right\} },
\eear
\ee
where $\mu$ stands for $M_0$ or $M$.
One can still choose a special direction for the ``external'' field $\vec{W}$:
$$\vec{W}= (0, 0, W).$$
Then \eq{eq:5.7} becomes:\footnote{In the following, unless stated
  otherwise, $\phi_{1,2,3}$ will refer to the three colour components of
  the remaining chiral superfield, $\phi$.} 
\be\label{eq:5.8}
4 W_1 W_2 \left(1 + \frac{\phi^2}{(p^2 + \mu)^2} \right)^{-2} 
\left[ 1 + \frac{\phi_1^2 + \phi_2^2 - \phi_3^2}{(p^2 + \mu)^2} \right].
\ee

Unfolding the $\dsla$ integral,  
$$\dbar p = \frac{1}{16\pi^2} \, p^2 \, dp^2 = \frac{1}{16\pi^2} \, d\tau
\, \tau,$$  
the integral to be effected in \eq{eq:5.4} takes the form 
\be\label{eq:5.9}
\bear
{\ds \frac{4W_1 W_2}{16\pi^2} \int_0^\infty \!\! d\tau \, \tau
\bigg[ \bigg\{ \frac{(\tau + M_0)^2 - \phi^2}{((\tau + M_0)^2 + \phi^2)^2} - 
\frac{(\tau + M)^2 - \phi^2}{((\tau + M)^2 + \phi^2)^2} \bigg\} }\\[0.35cm]
{\ds + 2(\phi_1^2 + \phi_2^2)\bigg\{ \frac{1}{((\tau + M_0)^2 + \phi^2)^2} 
- \frac{1}{((\tau + M)^2 + \phi^2)^2} \bigg\}\bigg].}
\eear
\ee
The terms in the first pair of curly brackets give: 
\be\label{eq:5.10}
\frac{2 W_1 W_2}{16\pi^2} \left[ \ln\left(\frac{M}{M_0}\right)^2 + \ln \left( \frac{1
  + \frac{\phi^2}{M^2}} {1 + \frac{\phi^2}{M_0^2}} \right) \right].
\ee
The terms in the second line of \eq{eq:5.9} are proportional to the 
projection of $\phivec$ in the direction orthogonal to that of $\vec{W}$ and give: 
\be\label{eq:5.11}
\frac{4W_1 W_2}{2 \cdot 16\pi^2} \left[ \frac{2(\phi_1^2 +
    \phi_2^2)}{M^2}\left (\frac{M}{\phi}\right) 
^3  \left( \arctan \left(\frac{\phi}{M}\right) - \frac{\phi}{M} \right) - ( M \to M_0)
  \right]. 
\ee
Together they give the effective potential term for $\phivec$, 
%%% (=\phivec_3)$,   
which can now be reduced.

\subsection{${\phi}$ reduction}
Now we can apply the methods of secs.\ \ref{sec:SUSY},\ref{sec:regu} to the
effective action we have obtained for $\phivec$.
We limit ourselves to the Gaussian approximation of the potential terms,
\eqs{eq:5.10}{eq:5.11}. %%%, \ie we replace their sum by
Adding the terms from $(\wh_1
    \leftrightarrow \wh_2)$, we obtain 
\be\label{eq:5.12}
\frac{4W_1 W_2}{16\pi^2} \left[ \ln\left(\frac{M}{M_0}\right)^2 +
  \left(\frac{\phi}{M}\right)^2  
- \frac{2}{3}\frac{\phi_1^2 + \phi_2^2}{M^2} \right],  
\ee
where the limit $M_0 \to \infty$ has been anticipated and  
$\phi / M_0$  terms have been discarded.

Thus, to reduce $\phivec$ and $\bar{\phivec}$, we have to 
start from the action:
\be\label{eq:5.13}
\bear
{\ds S_{eff}(\phivec, \bar{\phivec}) = \frac{1}{\gzero^2} \int 
\phibar \, \e^V \phi + \frac{M_0}{2} 
\int \phi^2 + \frac{1}{16} \frac{1}{4 \pi^2} \int \Big(\frac{\phi_1^2 + 
\phi_2^2}{3 M^2}+ \frac{\phi_3^2}{M^2}
\Big) W^2 }\\
{\ds + \frac{\overline{M}_0}{2}\int \phibar^2 + O(\phi^4 \mbox{ term }) + 
\mbox{ (irrelevant non holomorphic part) }.}
\eear
\ee
From \eq{eq:5.13} we can compute the contribution to $S_M$ proceeding 
exactly as before, \ie first
we will integrate out $\phihat \equiv \phibar \, \e^V$ and then we will 
apply the RG transformation 
$\frac{1}{2} M_0 \phi^2 \to \frac{1}{2} M \phi^2$.
For the configuration $\vec{V}= (0, 0, V)$ it is convenient to use the 
following linear 
combinations of the fields:\footnote{Remember that $\phi_1, \phi_2$ and 
$\phi_3$ all refer to the 
gauge indices of the third chiral field $\phivec (= \phivec_3)$.}
\bea\label{eq:5.14}
(\phi)_{1, 2} & \to & \phi_\pm = \frac{1}{\sqrt{2}}(\phi_1 \pm i \phi_2), \\ \nonumber 
(\phi)_3 & \to & (\phi)_3. 
\eea
This gives
\bea\label{eq:5.15}
\phibar_a(\e^V)_{(ab)}\phi_b & = & \phibar_+\e^V\phi_+ + \phibar_-\e^{-V}\phi_- +\phibar_3 \phi_3,
\nonumber \\
\phi_1^2 + \phi_2^2 & = & 2 \phi_+ \phi_-.
\eea
After integrating out $\phihat = \phibar \, \e^V$, we have the effective 
action:\footnote{Please note that in \eq{eq:5.16}, $\phivec_{1,2}$ refer to
  the two massive  
chiral fields, and this last linear term comes from the potential term in
the original$S_{\Nfour}$.} 
\be\label{eq:5.16}
\bear
{\ds \frac{1}{2} \sum_{I, J}^{+,-} \phistar_I \left( -p^2 + i \pi W
  \sigma_3 + \left(M_0 + \frac{S}{32\pi^2 \cdot 3 \cdot M^2} \right)
  \sigma_1 \right)_{(IJ)}\phistar_J }\\[0.35cm] 
{\ds + \frac{1}{2} \phistar_3 \left(-p^2 + M_0 + \frac{S}{32 \pi^2 \cdot M^2}
  \right)\phistar_3 + \frac{1}{\gzero^2}
(\phivec \cdot \phivec_1 \wedge \phivec_2), }
\eear
\ee
where we wrote $W^2 \equiv S$.
%%%The matrices $\sigma_3$ and $\sigma_1$ correspond to the terms of the
%%%action in \eq{eq:5.15}.

Applying the RG transformation ($M_0 \to M$), introducing the auxiliary fields 
($\phi_\pm', \phi_3'$) and diagonalizing, \eq{eq:5.16} is transformed to:
\be\label{eq:5.17}
\bear
{\ds \frac{1}{2}\int \dbar p d^2 \pi \bigg[ \phi_I' \left( -p^2 + i \pi W
    \sigma_3 +  
\left( M_0 + \frac{S}{32 \pi^2 \cdot 3 \cdot M^2} \right)\sigma_1
\right)_{IK} }\\[0.35cm] 
{\ds \times \left( -p^2 + i \pi W \sigma_3 +  
\left(M + \frac{S}{32 \pi^2 \cdot 3 \cdot M^2} \right)\sigma_1
\right)_{KJ}^{-1}\phi_J' }\\[0.35cm] 
{\ds + \phi_3' \left( -p^2 +  
M_0 + \frac{S}{32 \pi^2 \cdot 3 \cdot M^2} \right)
\left( -p^2 +  
M + \frac{S}{32 \pi^2 \cdot 3 \cdot M^2 }\right)^{-1} \phi_3'}\\[0.35cm]
{\ds + \mbox{ linear term in } (\phi_\pm', \phi_3') + (\phi_\pm, \phi_3) \mbox{ part } \Big],}
\eear
\ee
with $I,J,K = +,-$.

The linear term comes from the last term in \eq{eq:5.16} after the
diagonalization. 
Discarding it for the moment, we can effect the path integral $\int \D
\phi_\pm' \D \phi_3'$. 
After a Wick rotation, the relevant factor for $Z_{M_0}$ comes from the
bilinear term in   
\eq{eq:5.17}:
\be\label{eq:5.18}
\bear
{\ds \exp -\frac{1}{2}\int \dbar p d^2 \pi \tr \bigg\{\ln\Big(p^2 + i \pi W
  \sigma_3 + 
  \Big(M_0 + \frac{S}{32 \pi^2 \cdot 3 \cdot M^2}\Big)\sigma_1 \Big)
}\\[0.35cm] 
{\ds - \ln\Big(p^2 + i \pi W \sigma_3 +
\Big(M + \frac{S}{32 \pi^2 \cdot 3 \cdot M^2} \Big)\sigma_1 \Big) \bigg\}. }
\eear
\ee
Note that the integral over the ``neutral'' field $\phi_3'$ gives only a
vanishing contribution 
in our approximation scheme (cf. \sec{sec:regu}). Effecting the $\pi_1,
\pi_2$ integral in   
\eq{eq:5.18}, the exponent becomes:
\be\label{eq:5.19}
\frac{i W^2}{4} \int \dbar p \tr \bigg\{(p^2 + \M \sigma_1)^{-1}\sigma_3
(p^2 + \M \sigma_1)^{-1} \sigma_3 \bigg\}_{\M=M + \frac{S}{32\pi^2
    \cdot 3 \cdot M^2}} 
^{\M=M_0 + \frac{S}{32\pi^2 \cdot 3 \cdot M^2}}.
\ee
Effecting the integral over $\dbar p = \frac{\tau d\tau}{16 \pi^2}$ one
obtains: 
\be\label{eq:5.20}
\frac{i}{4} \frac{W^2}{16 \pi^2} \ln \left( \frac{\ds M + \frac{S}{32\pi^2
    \cdot 3 \cdot M^2}}{\ds M_0 +  
 \frac{S}{32\pi^2 \cdot 3 \cdot M^2}} \right)^2.
\ee
Thus the contribution \eq{eq:5.18} to $\Z_{M_0}$ becomes
\be\label{eq:5.21}
\exp \frac{i}{4} \frac{1}{16 \pi^2} \inte \left\{
\ln\left(\frac{M}{M_0}\right)+  
\ln \left( \frac{\ds 1 + \frac{S}{32\pi^2 \cdot 3 \cdot M^3}}{\ds 1 + 
\frac{S}{32\pi^2 \cdot 3 \cdot M_0 M^2}} \right) \right\} S,
\ee
where, as usual, $S\equiv W^2$.
To this one must add
\begin{description}
\item[a)] the residual constant term coming from the computation in
  \sec{sec:phi1phi2} [\cf \eq{eq:5.12}]:
$$
\exp \frac{i}{4} \frac{1}{16 \pi^2}\int \ln\left(\frac{M}{M_0}\right)^2 S;
$$
\item[b)] the gauge kinematic term in $S_{N=4}$:
$$
\exp \frac{i}{16}\int \left(\frac{1}{\gzero^2} + \frac{i \theta_0}{8 \pi^2}\right)S = 
\exp \frac{i}{8}\frac{1}{16 \pi ^2} \int \left(2 \ln\left( \frac{M_0}{\Lam_{\Nuno}} \right)^3 + i\theta_0\right)S.
$$
\end{description}
Putting together all the terms, one obtains
\be\label{eq:5.22}
\exp \frac{i}{8}\frac{1}{16 \pi ^2}\int 
\left[ 
2 \ln\left( \frac{M_0}{\Lam_{\Nuno}} \right)^3 +
2 \ln \left( \frac{\ds 1 + \frac{S}{32\pi^2 \cdot 3 \cdot M^3}}{\ds 1 + 
\frac{S}{32\pi^2 \cdot 3 \cdot M_0 M^2}} \right) + 
i \theta_0
\right] S.
\ee
Now choose $M$ so that $\frac{S^{1/3}}{M}\gg 1$ and consider the pure SYM limit 
$
M_0 \to \infty, \gzero \to 0$ with $\Lam_{\Nuno}$ fixed:
\be\label{eq:5.23}
\exp  \frac{i}{8}\frac{1}{16 \pi ^2}\inte \left[ 2\ln\left(
  \frac{S}{32\pi^2 \cdot 3 \cdot \Lam^3} \right)S + 
i\theta_0 S \right],
\ee
which gives the effective potential
\be\label{eq:5.24}
W_{eff}= \frac{1}{128\pi^2} \inte \left[2 \ln\Big(\frac{S}{3 \cdot \Lam^3
    \cdot 32 \pi^2}\Big)S +  i \theta_0 S \right].
\ee
\ceq{eq:5.24} is of the VY form \cite{VY}.

One can study the vacuum structure of $\Nuno$ SYM looking for the minima of
 $W_{eff}$
\be\label{eq:5.25}
\left(\frac{ \partial W_{eff} }{\partial S} \right)_{\bar{S}}=0.
\ee
Bearing in mind that $\theta_0$ is defined only up to $2k\pi, k = 0, \pm 1, \pm 2 \dots$, the
solution to \eq{eq:5.25} is:
\be\label{eq:5.26}
\frac{\bar{S}}{32\pi^2} = \frac{3}{\e} \left(\pm \e^{-\frac{i \theta_0}{2} }\right) \Lam^3.
\ee
One can now conclude that
\begin{description}
\item[i.] $\langle W^2 \rangle \not = 0 $
\item[ii.] the vacuum is two-fold degenerate.
\end{description}
For SU($N_c$) one expects, instead of \eq{eq:5.24}, 
\be\label{eq:5.27}
W_{eff} \propto  \inte \left[N_c S \ln\Big(\frac{S}{3 \cdot \Lam^3 \cdot 32 \pi^2}\Big) + 
i \theta_0 S \right],
\ee
which predicts a $N_c$-fold degenerate vacuum.

\subsection{Linear term}
In \eq{eq:5.17} we have left out the linear term in the auxiliary fields $\phi_\pm'$ and 
$\phi_3'$. The effect of such a term is to generate an additional four-point
vertex in the effective 
potential which is not strictly local. However, in the limit of low energy,
this vertex reduces to 
\be\label{eq:5.28}
\sim \frac{1}{2M_0}\int d^4x d^2 \theta \Big( \phivec_1 \wedge \phivec_2 \Big)^2. 
\ee
\ceq{eq:5.28} is potentially interesting because it breaks the original
``flavour'' symmetry  
of SU(2) among $(\phi_i)_{i=1}^{3}$ to SO(2) (the rotations around $\phi_3$). The effect of
such a term is negligible in the limit $M_0 \to \infty$. 

\section{Conclusions and problems}\label{Conc}

In this note we have presented an elementary QFT computation of the
glueball superpotential of the pure $\Nuno$ SYM model. 

In spite of the simplicity of our method, we have obtained the standard
Veneziano-Yankielowicz potential with minima of the ``right'' order of
magnitude and correct multiplicity structure for the model we have studied. 

As already noted in the introduction, the covariant super Feynman
graph technique we have adopted is perturbatively equivalent to the matrix
model approach \cite{dijk-vafa,zanon1}.

Only in the case that one applies directly the method of \cite{zanon1}, UV
divergences are likely to be run into, in  particular at the one-loop
level. In \cite{amb}, one indeed obtains the leading VY potential, which,
however, contains a large UV cutoff parameter.

On the other hand, if one attempts the same computation making use of
the MM correspondence \cite{dijk-vafa}, one might expect to get the same
result, but without encountering any UV divergences. Indeed, this appears
to be the result of the recent piece of work, \cite{HG}, where the leading
term is again given by the VY potential, but the UV cutoff is replaced by
physical finite parameters ($\sqrt[3]{m \Lam_{\Ndue}^2}$).

From \cite{HG}, the general attitude is somewhat incomprehensible of
limiting the application of the matrix model approach to the perturbative
corrections to the VY potential \cite{argurio}.

Our method, based on the use of covariant super Feynman graphs
\cite{zanon1}, is less likely to be troubled by UV divergences, even in the
computation of the leading term of the superpotential. This is due to the
incorporation of the ERG method and the corresponding Wilsonian action.
As a matter of fact, our method must be considered as the closest possible
QFT equivalent of the MM approach.

It is fairly straightforward to reproduce the result on the VY potential in
\cite{HG} within our method, although the model there considered is
different, being $\Nuno$ SYM with an additional chiral superfield in the
adjoint representation.

In detail, in the calculation presented in \sec{sec:Nuno}, one should
ascribe the $\phi (= \phi_3)$ field with the independent physical mass $m$,
instead of $M_0$, making it a physical degree of freedom. (This is
analogous to the case of $\Nfour$ SYM regularization of the $\Ndue$ model
\cite{afy}.) 

Repeating, step by step, the computations of \sec{sec:Nuno} with $\phivec$
having physical mass $m$, and RG-reducing $m$ to some other
value $m' \ll m$, we obtain, instead of \eq{eq:5.20},
\be\label{eq:6.2}
\frac{i}{4} \frac{W^2}{16 \pi^2} \ln \Big( \frac{m' + \frac{S}{32\pi^2
    \cdot 3 \cdot M^2}}{m +  
\frac{S}{32\pi^2 \cdot 3 \cdot M^2}} \Big)^2.
\ee 
One may get rid of the kinematical term by choosing
\be\label{eq:6.3}
\Big(\ln\Big(\frac{M}{M_0}\Big)^2 +
\frac{4\pi}{\gzero^2}\Big)_{M=\Lam_{\Ndue}}=0 \quad  
\mbox{ for G = SU(2),}
\ee
that is defining $\Lam_{\Ndue}$ to be the value at which the inverse gauge
coupling in the $\Ndue$ model vanishes.

Thus the effective potential is
\be
\frac{i}{4} \frac{W^2}{16 \pi^2}\ln \Big( \frac{m' + \frac{S}{32\pi^2 \cdot
    3 \cdot {\Lam_{\Ndue}}^2}} 
{m + \frac{S}{32\pi^2 \cdot 3 \cdot {\Lam_{\Ndue}}^2 }} \Big).
\ee
By choosing $m' \ll m$ and assuming $\frac{S}{32\pi^2 \cdot 3 \cdot m
  {\Lam_{\Ndue}}^2} \ll 1 $, the superpotential becomes proportional to
\be\label{eq:6.4}
\Big(2 S \ln \Big(\frac{S}{32\pi^2 \cdot 3 \cdot m{\Lam_{\Ndue}}^2} \Big) +
i \theta_0 \Big),
\ee
which is the special case of the result in \cite{HG} for $N_c = 2$. 

In order to obtain the glueball superpotential in $\Nuno$ SYM model, we
have made rather heavy use of ERG techniques, proposed in \cite{ah-m,afy} 

In the past, the VY term for $\Nuno$ SYM has been obtained as a one-loop
effect in the context of the correspondence between $\Ndue$ supersymmetric
sigma model in 2D and $\Nuno$ SYM in 4D, established through $T_2$
compactification.
The superpotential is supposed to be immune to volume effects in the
compactified space \cite{Agan}. In our approach, instead of dimensional
reduction, the ERG method produces convergent expressions directly in
four dimensions. 

On the other hand, as noted in \cite{ah-m}, our method is applicable only to
those models which can be obtained as mass deformation of $\Nfour$ SYM. A
little wider class of models can be dealt with by mass deforming the
$\Ndue$ SQCD with vanishing one-loop beta function, \ie with twice as many
flavours as colours. Even then, though, one cannot study any general SGFT,
whereas in the MM approach such restriction seems not to be present.

As it is, our method - which can be said to be the ``QFT version'' of MM
techniques - does not reach as yet  the same level of unified
understanding of SUSY GFT the string approach does \cite{Imeroni}.

\section*{Note added}

After the completion of the present work, the paper
``Veneziano-Yankielovicz superpotential terms in N=1 SUSY gauge theories''
by Gripaios and Wheater \cite{Gripaios} has come to our notice. 

In this work, the authors take a ``current algebra'' approach, \ie they
rely on the Konishi anomalous Ward-Takahashi identities (AWTI) \cite{konishi}
(see also \cite{argurio}), instead of the Feynman graph approach of
\cite{amb} and the present work. 

Out of the well-known Konishi-type AWTI,
the authors have succeeded in extracting the glueball superpotential (the
part relative to the pure glueball dynamics) as the residue after the
decoupling of the heavy fields, in the limit that the quark mass, $m$, and
the Higgs-induced gauge boson mass, $\sqrt{m/2\lam}$, become
infinite. In this respect, this piece of work is not dissimilar to ours.

Since the approach in \cite{Gripaios} depends only on the Konishi anomaly 
and the corresponding supercurrent divergence equation, rather than on the
microscopic Lagrangian (with its regularization problems), it has the
advantage of allowing for the computation of quantities of interest for a
wide class of gauge groups and hypermultiplet contents.

We believe this work supplies the important link for understanding        
gluball dynamics which has been mentioned in the introductory section of
the present work.

We thank the referee for drawing our attention to the existence of this
paper.   

\section*{Acknowledgments}

We would like to thank G.~Immirzi, H.~Itoyama, G.~C.~Rossi, S.~Sugimoto and
D.~Zanon for helpful discussions. Part of this work has been carried out
while KY was a guest at YITP, Kyoto and at the Department of Physics, the
University of Tokyo. It is a pleasure for him to thank K.~Kugo, M.~Ninomiya 
and K. Fujikawa for their warm hospitality.

%%%\bibliography{glueballjhep}
%%%\bibliographystyle{hunsrt}

\appendix

\section{The holomorphic and canonical coupling constants}

As stated in \sec{sec:regu}, throughout the paper we have been using the
so-called holomorphic form for the action, \eq{esseN4}, and the 
corresponding gauge coupling constant,\footnote{In this appendix its real
  part will be denoted by $g_h$.}\eq{eq:3.2}.

Its running, even in the $\Nuno$ SYM model, is exausted at one loop, 
the corresponding $\beta$ function being
$$
\beta(g_h) = - \frac{3 N_c}{16 \pi^2} \, g_h^3.
$$  

If the dinamically generated scale, $\Lam_{\Nuno}$, is defined 
as the value at which the inverse coupling constant vanishes, one gets
precisely [\cf \eq{lamnuno}]
$$
\Lam_{\Nuno} = M_0 \exp - \frac{8 \pi^2}{3 N_c \, g_h^2}.
$$

Going over to the more conventional canonical form is not completely
trivial, and requires the use of the Konishi anomaly \cite{konishi}. The
relation between the two coupling constants at a given scale $\mu$ is
\cite{ah-m}  
$$
\frac{1}{g_h^2(\mu)} = \frac{1}{g_c^2(\mu)} + \frac{N_c}{4 \pi^2} \ln g_c(\mu).
$$
($g_c$ only runs down to $\mu_0 = 8 \pi^2 \e \Lam /N_c$ \cite{carlino}.)

Substituting the above in the expression of $\Lam_{\Nuno}$ gives
$$
\Lam_{\Nuno} = \frac{M_0}{g_c^{2/3}(M_0)} \exp - \frac{8 \pi^2}{3 N_c \,
  g_c^2(M_0)},
$$
which has the correct dependence upon $g_c$ as expected from explicit
instanton calculations \cite{veneziano}.

\end{document}